\documentclass[conference]{IEEEtran}
\IEEEoverridecommandlockouts

\usepackage{cite}
\usepackage{amsmath,amssymb,amsfonts}
\usepackage{algorithmic}
\usepackage{graphicx}
\usepackage{textcomp}
\usepackage{xcolor}
\usepackage{hyperref}

\usepackage{float}
\usepackage{booktabs}
\usepackage{multirow}
\usepackage{kotex}

\def\BibTeX{{\rm B\kern-.05em{\sc i\kern-.025em b}\kern-.08em
    T\kern-.1667em\lower.7ex\hbox{E}\kern-.125emX}}
\begin{document}

\title{VoiceGuider: Enhancing Out-of-Domain Performance in Parameter-Efficient Speaker-Adaptive Text-to-Speech via Autoguidance
\\
 \thanks{* Corresponding Author}
%%%%%%%%%%%%%%%%%%%%%%%%%%%%%%%%%%
% acknowledgement
\thanks{This work was supported by Samsung Electronics Co., Ltd (IO240124-08661-01) and Korean Government (2022R1A3B1077720, 2022R1A5A708390811, RS-2022-II220959, BK21 Four Program, IITP-2024-RS-2024-00397085 \& RS-2021-II211343: AI Graduate School Program)}
%%%%%%%%%%%%%%%%%%%%%%%%%%%%%%%%%%
}

\author{\IEEEauthorblockN{Jiheum Yeom} 
\IEEEauthorblockA{\textit{Electrical and Computer Engineering} \\
\textit{Seoul National University}\\
Seoul, Republic of Korea \\
quilava1234@snu.ac.kr}
\and
\IEEEauthorblockN{Heeseung Kim} 
\IEEEauthorblockA{\textit{Electrical and Computer Engineering} \\
\textit{Seoul National University}\\
Seoul, Republic of Korea \\
gmltmd789@snu.ac.kr}
\and
\IEEEauthorblockN{Jooyoung Choi}
\IEEEauthorblockA{\textit{Electrical and Computer Engineering} \\
\textit{Seoul National University}\\
Seoul, Republic of Korea \\
jy\_choi@snu.ac.kr}
\and
\IEEEauthorblockN{Che Hyun Lee} 
\IEEEauthorblockA{\textit{Electrical and Computer Engineering} \\
\textit{Seoul National University}\\
Seoul, Republic of Korea \\
saga1214@snu.ac.kr}
\and
\IEEEauthorblockN{Nohil Park}
\IEEEauthorblockA{\textit{Electrical and Computer Engineering} \\
\textit{Seoul National University}\\
Seoul, Republic of Korea \\
pnoil2588@snu.ac.kr}
\and
\IEEEauthorblockN{Sungroh Yoon$^{*}$}
\IEEEauthorblockA{\textit{ECE, AIIS, ASRI, INMC, ISRC, and IPAI} \\
\textit{Seoul National University}\\
Seoul, Republic of Korea \\
sryoon@snu.ac.kr}
}

\maketitle

\begin{abstract}
When applying parameter-efficient finetuning via LoRA onto speaker adaptive text-to-speech models, adaptation performance may decline compared to full-finetuned counterparts, especially for out-of-domain speakers. Here, we propose VoiceGuider, a parameter-efficient speaker adaptive text-to-speech system reinforced with autoguidance to enhance the speaker adaptation performance, reducing the gap against full-finetuned models. We carefully explore various ways of strengthening autoguidance, ultimately finding the optimal strategy. VoiceGuider as a result shows robust adaptation performance especially on extreme out-of-domain speech data. We provide audible samples in our demo page.
\end{abstract}

\begin{IEEEkeywords}
text-to-speech, speaker adaptive text-to-speech, diffusion, autoguidance, parameter-efficient text-to-speech
\end{IEEEkeywords}

\section{Introduction}

While deep generative models, such as diffusion models \cite{pmlr-v37-sohl-dickstein15, DDPM, song2021scorebased}, have consistently demonstrated impactful advancements \cite{NEURIPS2021_49ad23d1, ho2021classifierfree, pmlr-v202-song23a}, text-to-speech (TTS) models have mirrored these achievements, advancing significantly in both quality and efficiency \cite{popov2021grad, pmlr-v162-kim22d}. Speaker-adaptive TTS, which aims to synthesize speech from speakers unseen during pretraining, reflects this trend of innovation. These advancements are primarily categorized into two approaches: zero-shot adaptation \cite{Guided-TTS2, pmlr-v162-casanova22a, VALL-E, P-Flow, NaturalSpeech2, le2023voicebox} and few-shot adaptation \cite{Guided-TTS2, chen2021adaspeech, adaspeech2, hsieh23_interspeech, UnitSpeech}. While zero-shot adaptation eliminates the need for additional training, it typically requires substantial investments in training resources, model size, and dataset comprehensiveness.

Conversely, few-shot adaptation models are generally recognized to outperform their zero-shot counterparts while significantly reducing training costs \cite{Guided-TTS2, UnitSpeech}. Studies have demonstrated personalization using several minutes to as little as 5$\sim$10 seconds of reference data per target speaker \cite{chen2021adaspeech, adaspeech2, Guided-TTS2, UnitSpeech}, where the latter is also acknowledged as one-shot adaptation model. Furthermore, some researches aim to achieve speaker adaptation with minimal parameter increments and performance drops \cite{adaspeech2, hsieh23_interspeech}, utilizing methods such as Low-Rank Adaptation (LoRA) \cite{LoRA}. Among these, VoiceTailor \cite{VoiceTailor} successfully combines a pretrained diffusion-based decoder with LoRA to establish an efficient framework for training plug-in adapters for one-shot speaker-adaptive TTS.

While various parameter-efficient one-shot TTS models demonstrate strong adaptation performance, their evaluations are mostly conducted on in-domain data similar to datasets used for pretraining, often comprising well-constrained speech data such as audiobook recordings \cite{librispeech, ljspeech, zen19_interspeech}. In real-world scenarios, demands for handling in-the-wild reference data grow, yet obtaining sufficient quantities of such data for pretraining is typically challenging. Therefore, achieving robustness to out-of-domain (OoD) data during speaker adaptation, particularly with in-the-wild recordings, remains a critical challenge. Although the previous baseline, VoiceTailor, shows reasonable speaker adaptation performance on in-domain data comparable to full-finetuning, significant performance degradation is observed when adapting to OoD data that substantially deviates from the pretraining distribution.

\begin{figure*}[h]
    \centering
    \includegraphics[width=0.80\linewidth]{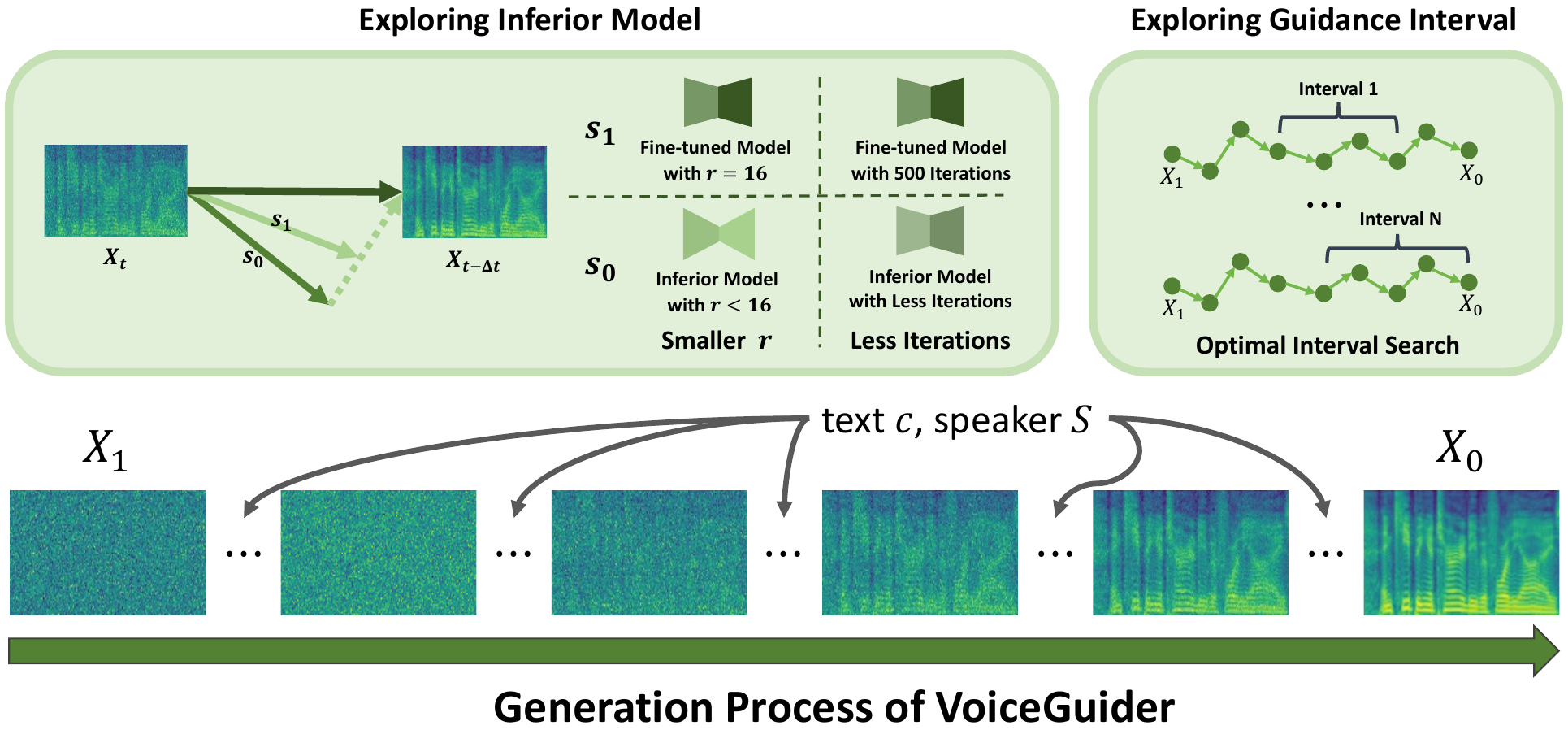}
    \caption{\textbf{Overview of VoiceGuider}. {VoiceGuider is a parameter-efficient one-shot TTS method for out-of-domain speakers. To eliminate errors caused by the parameter-efficient adapter, VoiceGuider enhances the prediction of the adapted model $s_1$ by pushing away from the predictions of the inferior model $s_0$. We explore two key ingredients of VoiceGuider: the method for obtaining the inferior model (left) and the guidance interval (right).}}
    \label{fig1}
\end{figure*}

In this work, we propose VoiceGuider, a parameter-efficient one-shot TTS model designed to robustly adapt to OoD reference data. We utilize VoiceTailor \cite{VoiceTailor}, a parameter-efficient one-shot TTS model, as our backbone model. By incorporating autoguidance \cite{autoguidance}, which enhances conditioning by guiding generation with a degraded model, we aim to improve speaker adaptation performance for reference data significantly far from the training distribution, such as `in-the-wild' recordings. In addition to exploring degraded model candidates proposed in \cite{autoguidance}, we analyze various candidates tailored to parameter-efficient models to identify the optimal autoguidance strategy for VoiceTailor. Furthermore, by examining the integration of autoguidance at different stages of the generation process, VoiceGuider effectively incorporates target speaker information, thereby enhancing speaker adaptation for OoD data.

We demonstrate that VoiceGuider achieves comparable performance to full-finetuning baselines on GigaSpeech \cite{chen21o_interspeech}, a dataset including in-the-wild data, while surpassing parameter-efficient one-shot baseline. Additionally, we assess the effectiveness of various guidance techniques for parameter-efficient one-shot TTS by comparing different degraded model candidates proposed for autoguidance. The robustness of our model across various in-the-wild data samples, and the samples used in our evaluations, is showcased on our demo page\footnote{{Demo: \href{https://voiceguider.github.io/}{https://voiceguider.github.io/}}}.

\section{Methodology}

In this section, we explain the background, baseline model, and the methods used to alleviate OoD performance degradation problem in parameter-efficient speaker adaptive TTS. 

\subsection{Background}
\textbf{Denoising Diffusion.}
\label{method:diffusion}
Diffusion models \cite{DDPM} are generative models that add Gaussian noise to data in multiple steps and learn a denoising process to generate data. In the case of speaker-adaptive TTS, given a text embedding $c$ and a speaker embedding $S$, the diffusion models are trained to recover the noise $\epsilon_t$ added to the noisy mel-spectrogram $X_t$ using the following objective function:
\begin{align}   
    \label{eq:objective}        
        L&(\theta)={\mathbb{E}_{t,X_0,\epsilon_t}[\lVert\sqrt{1 - \lambda_t}s_\theta(X_t|c,S)+\epsilon_t\rVert_2^2]},
\end{align}
where $s_\theta$ is a diffusion model and $\lambda_t$ is a predefined noise schedule of GradTTS~\cite{popov2021grad} and $t\sim[0,1]$ indicates noise level.

\textbf{Diffusion Guidance.}
Recently, diffusion models employ classifier-free guidance (CFG)~\cite{ho2021classifierfree} to improve sample quality and the likelihood of given conditions. At each generation step, CFG modifies the model's prediction with an extrapolation between two predictions:
\begin{align}   
    \label{eq:cfg}        
        \hat{s}_\gamma(X_t|c,S)=s_\theta(X_t|c,S)+\gamma(s_\theta(X_t|c,S)-s_\theta(X_t|c,\emptyset)),
\end{align}
where $\gamma$ is a guidance scale. 
CFG pushes away the unconditional distribution to \textit{avoid} undesired speaker, thereby increasing the likelihood of speaker $S$.

\subsection{VoiceTailor: Parameter-Efficient Baseline}\label{AA}

Given a pretrained diffusion model, VoiceTailor~\cite{VoiceTailor} trains a parameter-efficient adapter, namely LoRA~\cite{LoRA}, to adapt to a new speaker using \eqref{eq:objective}. For each linear layer $W_0$ in attention modules of the pretrained diffusion model, VoiceTailor trains only a new matrix $\Delta W = B A$ where $B \in \mathbb{R}^{d \times r}$ and $A \in \mathbb{R}^{r \times k}$, thus $W = W_0+\alpha\cdot B A$ with scale $\alpha$. By choosing small rank $r$, VoiceTailor trains only 0.25\% of the whole parameters. Despite the small number of parameters, VoiceTailor achieves comparable speaker adaptation performance to the full-finetuning baseline, UnitSpeech~\cite{UnitSpeech}.

However, while VoiceTailor demonstrates strong performance for in-domain speakers closer to the pretrained domain, we observe that it falls behind Unitspeech for OoD speakers.
\begin{table}[htbp]
    \centering
    \caption{Dataset specifications. Here, the term \textbf{Pretrain} and \textbf{In-the-wild} refers to  whether the dataset was used for pretraining and contains in-the-wild OoD data, respectively.}
    \label{tab:dataset}
    \begin{tabular}{c|ccc}
        \toprule
        \textbf{Dataset}     & \textbf{Pretrain}   & \textbf{OoD}   & \textbf{In-the-wild}\\
        \midrule
        LibriTTS\cite{zen19_interspeech}          & $\checkmark$    & $\times$     & $\times$                      \\ 
        VCTK\cite{Yamagishi2019-pe}              & $\times$       &$\checkmark$      & $\times$                      \\ 
        GigaSpeech\cite{chen21o_interspeech}    & $\times$       &$\checkmark$      & $\checkmark$            \\ 
        \bottomrule
    \end{tabular}
\vskip -0.2in
\end{table}

\subsection{VoiceGuider: Eliminating LoRA Error with Autoguidance}
\label{method:voiceguider}

We argue that the limited capacity of the LoRA in VoiceTailor~\cite{VoiceTailor} results in prediction errors for OoD speakers, and these errors are amplified by the iterative generation process of the diffusion model, leading to subpar performance compared to full-finetuning. Given that CFG effectively steers away from undesired samples, we propose VoiceGuider, a diffusion guidance to \textit{eliminate LoRA errors} in speaker-adaptive TTS.

\textbf{Autoguidance.} A concurrent study on diffusion guidance ~\cite{autoguidance} shows that a strong conditional model and an unconditional model share correlated errors. The study empirically shows that since the unconditional model’s error is more over-emphasized, the extrapolation in CFG \textit{eliminates} this error. \cite{autoguidance} further proposes autoguidance, which guides a strong model $s_1(X_t|c,S)$ with an inferior model $s_0(X_t|c,S)$ instead of the unconditional one $s_1(X_t|c,\emptyset)$, and demonstrates its effectiveness in conditional image synthesis. In our case of speaker-adaptive TTS, we combine CFG and autoguidance to amplify speaker likelihood and reduce LoRA errors:
\begin{align}   
    \label{eq:autoguidance}        
        \hat{s}_\gamma(X_t|c,S)=s_1(X_t|c,S)+&\underbrace{\gamma_S(s_1(X_t|c,S)-s_1(X_t|c,\emptyset))}_{\text{Speaker likelihood}~\uparrow} \notag \\
        +&\underbrace{\gamma_a(s_1(X_t|c,S)-s_0(X_t|c,S))}_{\text{LoRA error}~\downarrow},
\end{align}
where $\gamma_S$ and $\gamma_a$ denote scales of CFG and autoguidance.
Furthermore, as illustrated in Fig.~\ref{fig1}, we aim to identify the optimal autoguidance strategy for speaker-adaptive TTS from the following candidates of the inferior model $s_0$:
\begin{itemize}
\item \textbf{Shorter training time:} We construct $s_0$ using a model that consumed a shorter training time than $s_1$. This is the simplest approach to obtain inferior model, as it can be derived from the intermediate checkpoints of $s_1$.
\item \textbf{Smaller LoRA rank:} We construct $s_0$ using a smaller LoRA rank size than $s_1$. Our baseline, VoiceTailor, demonstrated that smaller ranks, such as 2 and 4, are inferior compared to a rank size of 16.
\end{itemize}

\textbf{Limited guidance interval.} Along with exploring inferior models, we later demonstrate that autoguidance can be detrimental in certain interval of the generation process. Specifically, at high noise levels, the guidance may blindly push away from the data distribution, leading to mode dropping and degraded sample quality~\cite{Interval}. We further demonstrate that guidance in the range of $t \in [0.6, 1.0]$ is detrimental, and disabling both classifier-free guidance and autoguidance in this range reduces errors and achieves higher speaker similarity. Thus, both guidance scales $\gamma_S$ and $\gamma_a$ can be written as:

\begin{align}   
    \label{limited_interval}        
        \gamma(t) = \begin{cases} 
            \gamma & \text{if } t \in (t_{\text{lo}}, t_{\text{hi}}] \\
            0 & \text{otherwise}.
        \end{cases}
\end{align}

\begin{table}[htbp]
    \centering
    \caption{SECS results for problem statement verification results.}
    \label{tab:table1}
    \begin{tabular}{c|ccc}
        \toprule
        \textbf{Method}    & \textbf{LibriTTS}    & \textbf{VCTK}     & \textbf{GigaSpeech}\\
        \midrule                    
        Full-finetuned \cite{UnitSpeech}                & $0.948$     & $0.873$          & $0.902$                      \\ 
        LoRA-tuned \cite{VoiceTailor}          & $0.940$    & $0.854$          & $0.875$            \\ 
        \midrule
        Performance Gap          & $0.008$    & $0.019$          & \textbf{$0.027$}            \\ 
        \bottomrule
    \end{tabular}
\end{table}

In the disabled interval, the inferior model $s_0$  in autoguidance can be seen as equivalent to the finetuned model $s_1$.

\section{Experiments and Results}

\subsection{Experimental Setup}
\subsubsection{Datasets}

To first validate the performance degradation of parameter-efficient one-shot TTS in OoD reference data, we use three different TTS datasets. Each represents the in-domain dataset utilized during pretraining, the OoD dataset, and the extreme OoD dataset further apart from the in-domain such as `in-the-wild' data. Details of the datasets are specified in Table \ref{tab:dataset}.

As shown, we pretrain our model with LibriTTS, thus positioned as the in-domain dataset. VCTK is not used for training, thus becomes the OoD dataset. GigaSpeech is not used for pretraining and includes in-the-wild speech data, becoming the extreme OoD data. After using these three datasets to validate the degradation problem, we use only the extreme OoD data, GigaSpeech, for the rest of the experiments.

\subsubsection{Training details}

Our main model builds upon the structure of VoiceTailor, and is finetuned for 500 iterations on a single NVIDIA A40 GPU. All finetuning is processed with Adam optimizer\cite{Adam} at learning rate of $10^{-4}$. For LoRA settings, we use rank $r$ of 16 and scaling factor $\alpha$ of 8, equivalent to VoiceTailor. The inferior model for autoguidance is trained with the same hyperparameters of the main model except for the varying factors specified in Section \ref{results:ablation}.

\subsubsection{Evaluation settings}
\label{results:experimental setup:evaluation settings}
We use 50 sentences from the test set of each dataset for evaluation. Character error rate (CER) and speaker encoder cosine similarity (SECS) are evaluated each for pronunciation and speaker adaptation quality, using the CTC-based conformer\cite{gulati20_interspeech} of the NEMO toolkit \cite{kuchaiev2019nemo} and speaker encoder of Resemblyzer\cite{resemblyzer} package, respectively. For subjective evaluation, we evaluate with 5-scale mean opinion score (MOS) for naturalness and human preference tests asking which model sample resembles the reference more, regarding speaker adaptation quality compared to baseline.

For inferior model of VoiceGuider, we set rank $r=1$, scaling factor $\alpha=8$, and stop finetuning at 100 iterations, while finetuned model $s_1$ follows VoiceTailor's hyperparameters. For inference, we set both autoguidance scale ${\gamma_a}$ and CFG scale ${\gamma_S}$ to 1.0, and set guidance intervals $[t_{\text{lo}}, t_{\text{hi}}]$ to $[0.1, 0.6]$.

\begin{table}[htbp]
\centering
\caption{Model comparsion of VoiceGuider against various baselines. Bold text indicates the winning side with statisical significance (wilcoxon signed-rank test with $p<0.05$). }
\label{tab:table2}
\begin{tabular}{l|c|cc}
\toprule
\multirow{2}{*}{\textbf{Method}} & \multicolumn{1}{c|}{\textbf{Speaker Similarity}}                                                                             & \multicolumn{1}{c}{\multirow{2}{*}{\textbf{MOS}}} & \multicolumn{1}{c}{\multirow{2}{*}{\textbf{CER(\%)}}} \\ \cline{2-2}
                                 & \multicolumn{1}{c|}{\textbf{\textit{win} / \textit{\textbf{tie}}} / \textit{\textbf{loss}}} & \multicolumn{1}{c}{}                              & \multicolumn{1}{c}{}                                   \\ \midrule
Ground Truth                     & 45.6\% / 4.2\% / 50.2\%                                                                                                      & 3.92 $\pm$ 0.10                                   & 2.31\%                                                 \\ \midrule
XTTS \textit{v2} \cite{casanova24_interspeech}                         & 50.7\% / 4.0\% / 45.3\%                                                                                                      & 3.85 $\pm$ 0.10                                   & 1.53\%                                                 \\
CosyVoice \cite{du2024cosyvoicescalablemultilingualzeroshot}                       & 48.0\% / 5.1\% / 46.9\%                                                                                                      & 4.04 $\pm$ 0.09                                   & 4.24\%                                                 \\ \midrule
UnitSpeech \cite{UnitSpeech}                      & 41.6\% / 11.6\% / 46.9\%                                                                                                     & 4.10 $\pm$ 0.09                                   & 1.53\%                                                 \\
VoiceTailor \cite{VoiceTailor}                     & \textbf{51.1\%} / 9.6\% / 39.3\%                                                                                                      & 4.01 $\pm$ 0.09                                   & 1.56\%                                                 \\
VoiceGuider                      & $-$                                                                                                              & 4.00 $\pm$ 0.09                                   & 1.72\%         \\ \bottomrule                                   
\end{tabular} 
\vskip -0.2in    
\end{table}

For baselines, we use UnitSpeech for the full-finetuned adaptation model and VoiceTailor for the LoRA-tuned baseline. All model outputs are reconstructed with BigVGAN\cite{lee2023bigvgan}, using the official checkpoint. When conducting subjective evaluation and preference tests, we additionally deploy open-source pretrained zero-shot adaptation models of XTTS \textit{v2}\cite{casanova24_interspeech} and CosyVoice\cite{du2024cosyvoicescalablemultilingualzeroshot} to compare against. During evaluation, all samples are normalized to the same volume of -27dB and resampled to 16kHz for a fair comparison.

\subsection{Problem Statement Verification}

To verify the performance degradation issue on OoD speakers, we compare SECS values of full-finetuned model to LoRA-tuned model on the three datasets of in-domain, OoD, and in-the-wild OoD. 
Table \ref{tab:table1} shows the results on our experiments. The performance gap increases as we move further away in data domain, thus validating our problem statement.

\subsection{Model Comparison}

With the degradation problem verified, we continue on with main results of VoiceGuider, identifying whether our model succeeds in alleviating the issue.
Table \ref{tab:table2} shows the results of VoiceGuider compared to baseline models on GigaSpeech. Since GigaSpeech is consisted heavily of in-the-wild data, it is challenging to verify naturalness through generated results, as can be inferred from ground truth showing low MOS. VoiceGuider does show MOS equivalent to VoiceTailor, which indicates that VoiceGuider retains naturalness. For preference tests on speaker similarity, VoiceGuider gains the upperhand to VoiceTailor with a $p$-value of 0.009, implying superior speaker adaptation on in-the-wild OoD data. Additionally, VoiceGuider also shows comparable preference to the two zero-shot models, proving comparable performance to models which use heavy amounts of data for pretraining (27k hours for XTTS \textit{v2}, 172k hours for CosyVoice).

\subsection{Ablation Studies}
\label{results:ablation}

\begin{table}[h]
    \centering
    \caption{CER, SECS results on variations of VoiceGuider. Full-finetuning implies results of applying the optimal autoguidance of VoiceGuider into full-finetuned model. Bold values indicate hyperparameters used for our final model.}
    \label{tab:table3}
    \begin{tabular}{cc|cc}
        \toprule
        \multicolumn{2}{c|}{\multirow{2}{*}{}} & \multicolumn{2}{c}{\textbf{Text-to-Speech}} \\
        \cmidrule(lr){3-4} 
        \multicolumn{2}{c|}{}                  & \textbf{CER(\%)} $(\downarrow)$        & \textbf{SECS} $(\uparrow)$ \\
        \midrule
        \multicolumn{1}{c|}{\multirow{6}{*}{\# of Iterations}}                 & 0         & 7.98     & 0.847         \\ 
        \multicolumn{1}{c|}{}                                                           & \textbf{100}       & 1.71     & 0.891         \\ 
        \multicolumn{1}{c|}{}                                                           & 200       & 1.78     & 0.883         \\ 
        \multicolumn{1}{c|}{}                                                           & 300       & 1.78     & 0.879         \\ 
        \multicolumn{1}{c|}{}                                                           & 400       & 1.67     & 0.875         \\ 
        \multicolumn{1}{c|}{}                                                           & 500       & 1.69     & 0.872         \\ 
        \midrule
        \multicolumn{1}{c|}{\multirow{4}{*}{Rank $r$}}                             & \textbf{1}         & 1.71     & 0.891         \\ 
        \multicolumn{1}{c|}{}                                                           & 2         & 1.83     & 0.891         \\ 
        \multicolumn{1}{c|}{}                                                           & 4         & 1.88     & 0.890         \\ 
        \multicolumn{1}{c|}{}                                                           & 8         & 1.69     & 0.891         \\ 
        \midrule
        \multicolumn{1}{c|}{\multirow{5}{*}{\begin{tabular}[c]{@{}c@{}}Autoguidance\\ Scale $\gamma_a$\end{tabular}}}      & 0.00      & 1.63     & 0.877         \\ 
        \multicolumn{1}{c|}{}                                                           & 0.33      & 1.61     & 0.886         \\ 
        \multicolumn{1}{c|}{}                                                           & 0.66      & 1.69     & 0.890         \\ 
        \multicolumn{1}{c|}{}                                                           & \textbf{1.00}      & 1.71     & 0.891         \\
        \multicolumn{1}{c|}{}                                                           & 1.33      & 2.00     & 0.892         \\         
        \midrule
        \multicolumn{1}{c|}{\multirow{5}{*}{\begin{tabular}[c]{@{}l@{}}Interval Upper $t_{hi}$\\ \end{tabular}}}      & 0.9      & 1.94     & 0.893         \\ 
        \multicolumn{1}{c|}{}                                                           & 0.8      & 1.84     & 0.892         \\ 
        \multicolumn{1}{c|}{}                                                           & 0.7      & 1.73     & 0.897         \\ 
        \multicolumn{1}{c|}{}                                                           & \textbf{0.6}      & 1.72     & 0.902         \\
        \multicolumn{1}{c|}{}                                                           & 0.5      & 1.92     & 0.903         \\
        \midrule
        \multicolumn{1}{c|}{\multirow{5}{*}{\begin{tabular}[c]{@{}l@{}}Interval Lower $t_{lo}$\\ \end{tabular}}}      & 0.5      & 2.09     & 0.901         \\ 
        \multicolumn{1}{c|}{}                                                           & 0.4      & 2.59     & 0.898         \\ 
        \multicolumn{1}{c|}{}                                                           & 0.3      & 1.68     & 0.900         \\ 
        \multicolumn{1}{c|}{}                                                           & 0.2      & 1.88     & 0.901         \\
        \multicolumn{1}{c|}{}                                                           & \textbf{0.1}      & 1.72     & 0.902         \\
        \midrule
        \multicolumn{1}{c|}{\multirow{1}{*}{\begin{tabular}[c]{@{}l@{}}Full-finetuning\\ \end{tabular}}}      & -      & 1.89     & 0.904         \\ 
        \bottomrule
    \end{tabular}
\end{table}

Additionally, we conduct ablation experiments on various factors that may affect autoguidance used in VoiceGuider in Table \ref{tab:table3}. Among the hyperparameters of autoguidance explained in Section \ref{results:experimental setup:evaluation settings}, we experiment on different values of rank $r$ and number of finetuning iterations of the inferior model, analyzing on what setting of inferior model leads to the optimal results. We also perform ablation studies on the effects of autoguidance scale $\gamma_a$. We found best results at 100 iterations in terms of training iterations. Inferior models with smaller LoRA rank, on the other hand, showed no meaningful variance. Meanwhile, increasing the intensity of guidance through different gradient scale values, leads to increase of both CER and SECS. However, while SECS value converges at a certain saturation point, CER value continues to escalate.

We conducted tests not only on how to construct the weak model, but also on whether the guidance truly provides only positive influence across the whole generation process by controlling the upper boundary $t_{hi}$ and lower boundary $t_{lo}$ of the guidance interval. While the overall adaptation performance increases as the upper bound tightens, increasing the lower bound rather inflicts degradation of CER. When we applied the guidance methods of VoiceGuider onto the full-finetuned model, only minor amounts of performance increase were found, whereas VoiceGuider showed overall performance boosts, showing SECS equivalent to that of the full-finetuned model. These results indicate that VoiceGuider eliminates LoRA errors, reaching the performance of full-finetuning.

\section{Conclusion}

We proposed VoiceGuider, a parameter-efficient speaker adaptive TTS model with robust adaptation performance on in-the-wild OoD speakers. We first verified the performance degradation issue of LoRA-tuned models with OoD in speaker adaptive TTS experimentally. VoiceGuider, reinforced with autoguidance, showed enhanced performance on OoD data, successfully alleviating the performance gap to full-finetuned counterparts. Based on our results, we expect VoiceGuider to provide effective ways of building efficient personalization TTS models for broad ranges of real-world speakers.

\bibliographystyle{IEEEtran}
\bibliography{main}

\begin{thebibliography}{10}
\providecommand{\url}[1]{#1}
\csname url@samestyle\endcsname
\providecommand{\newblock}{\relax}
\providecommand{\bibinfo}[2]{#2}
\providecommand{\BIBentrySTDinterwordspacing}{\spaceskip=0pt\relax}
\providecommand{\BIBentryALTinterwordstretchfactor}{4}
\providecommand{\BIBentryALTinterwordspacing}{\spaceskip=\fontdimen2\font plus
\BIBentryALTinterwordstretchfactor\fontdimen3\font minus \fontdimen4\font\relax}
\providecommand{\BIBforeignlanguage}[2]{{%
\expandafter\ifx\csname l@#1\endcsname\relax
\typeout{** WARNING: IEEEtran.bst: No hyphenation pattern has been}%
\typeout{** loaded for the language `#1'. Using the pattern for}%
\typeout{** the default language instead.}%
\else
\language=\csname l@#1\endcsname
\fi
#2}}
\providecommand{\BIBdecl}{\relax}
\BIBdecl

\bibitem{pmlr-v37-sohl-dickstein15}
J.~Sohl-Dickstein, E.~Weiss, N.~Maheswaranathan, and S.~Ganguli, ``Deep unsupervised learning using nonequilibrium thermodynamics,'' in \emph{Proceedings of the 32nd International Conference on Machine Learning}, ser. Proceedings of Machine Learning Research, F.~Bach and D.~Blei, Eds., vol.~37.\hskip 1em plus 0.5em minus 0.4em\relax Lille, France: PMLR, 07--09 Jul 2015, pp. 2256--2265.

\bibitem{DDPM}
J.~Ho, A.~Jain, and P.~Abbeel, ``Denoising diffusion probabilistic models,'' in \emph{Advances in Neural Information Processing Systems}, H.~Larochelle, M.~Ranzato, R.~Hadsell, M.~Balcan, and H.~Lin, Eds., vol.~33.\hskip 1em plus 0.5em minus 0.4em\relax Curran Associates, Inc., 2020, pp. 6840--6851.

\bibitem{song2021scorebased}
Y.~Song, J.~Sohl-Dickstein, D.~P. Kingma, A.~Kumar, S.~Ermon, and B.~Poole, ``Score-based generative modeling through stochastic differential equations,'' in \emph{International Conference on Learning Representations}, 2021.

\bibitem{NEURIPS2021_49ad23d1}
P.~Dhariwal and A.~Nichol, ``Diffusion models beat gans on image synthesis,'' in \emph{Advances in Neural Information Processing Systems}, M.~Ranzato, A.~Beygelzimer, Y.~Dauphin, P.~Liang, and J.~W. Vaughan, Eds., vol.~34.\hskip 1em plus 0.5em minus 0.4em\relax Curran Associates, Inc., 2021, pp. 8780--8794.

\bibitem{ho2021classifierfree}
J.~Ho and T.~Salimans, ``Classifier-free diffusion guidance,'' in \emph{NeurIPS 2021 Workshop on Deep Generative Models and Downstream Applications}, 2021.

\bibitem{pmlr-v202-song23a}
Y.~Song, P.~Dhariwal, M.~Chen, and I.~Sutskever, ``Consistency models,'' in \emph{Proceedings of the 40th International Conference on Machine Learning}, ser. Proceedings of Machine Learning Research, A.~Krause, E.~Brunskill, K.~Cho, B.~Engelhardt, S.~Sabato, and J.~Scarlett, Eds., vol. 202.\hskip 1em plus 0.5em minus 0.4em\relax PMLR, 23--29 Jul 2023, pp. 32\,211--32\,252.

\bibitem{popov2021grad}
V.~Popov, I.~Vovk, V.~Gogoryan, T.~Sadekova, and M.~Kudinov, ``Grad-tts: A diffusion probabilistic model for text-to-speech,'' in \emph{International Conference on Machine Learning}.\hskip 1em plus 0.5em minus 0.4em\relax PMLR, 2021, pp. 8599--8608.

\bibitem{pmlr-v162-kim22d}
H.~Kim, S.~Kim, and S.~Yoon, ``Guided-{TTS}: A diffusion model for text-to-speech via classifier guidance,'' in \emph{Proceedings of the 39th International Conference on Machine Learning}, ser. Proceedings of Machine Learning Research, K.~Chaudhuri, S.~Jegelka, L.~Song, C.~Szepesvari, G.~Niu, and S.~Sabato, Eds., vol. 162.\hskip 1em plus 0.5em minus 0.4em\relax PMLR, 17--23 Jul 2022, pp. 11\,119--11\,133.

\bibitem{Guided-TTS2}
S.~Kim, H.~Kim, and S.~Yoon, ``Guided-tts 2: A diffusion model for high-quality adaptive text-to-speech with untranscribed data,'' \emph{arXiv preprint arXiv:2205.15370}, 2022.

\bibitem{pmlr-v162-casanova22a}
E.~Casanova, J.~Weber, C.~D. Shulby, A.~C. Junior, E.~G{\"o}lge, and M.~A. Ponti, ``{Y}our{TTS}: Towards zero-shot multi-speaker {TTS} and zero-shot voice conversion for everyone,'' in \emph{Proceedings of the 39th International Conference on Machine Learning}, ser. Proceedings of Machine Learning Research, K.~Chaudhuri, S.~Jegelka, L.~Song, C.~Szepesvari, G.~Niu, and S.~Sabato, Eds., vol. 162.\hskip 1em plus 0.5em minus 0.4em\relax PMLR, 17--23 Jul 2022, pp. 2709--2720.

\bibitem{VALL-E}
C.~Wang, S.~Chen, Y.~Wu, Z.-H. Zhang, L.~Zhou, S.~Liu, Z.~Chen, Y.~Liu, H.~Wang, J.~Li, L.~He, S.~Zhao, and F.~Wei, ``Neural codec language models are zero-shot text to speech synthesizers,'' \emph{ArXiv}, vol. abs/2301.02111, 2023.

\bibitem{P-Flow}
S.~Kim, K.~J. Shih, R.~Badlani, J.~F. Santos, E.~Bakhturina, M.~T. Desta, R.~Valle, S.~Yoon, and B.~Catanzaro, ``P-flow: A fast and data-efficient zero-shot {TTS} through speech prompting,'' in \emph{Thirty-seventh Conference on Neural Information Processing Systems}, 2023.

\bibitem{NaturalSpeech2}
K.~Shen, Z.~Ju, X.~Tan, E.~Liu, Y.~Leng, L.~He, T.~Qin, sheng zhao, and J.~Bian, ``Naturalspeech 2: Latent diffusion models are natural and zero-shot speech and singing synthesizers,'' in \emph{The Twelfth International Conference on Learning Representations}, 2024.

\bibitem{le2023voicebox}
M.~Le, A.~Vyas, B.~Shi, B.~Karrer, L.~Sari, R.~Moritz, M.~Williamson, V.~Manohar, Y.~Adi, J.~Mahadeokar, and W.-N. Hsu, ``Voicebox: Text-guided multilingual universal speech generation at scale,'' in \emph{Thirty-seventh Conference on Neural Information Processing Systems}, 2023.

\bibitem{chen2021adaspeech}
M.~Chen, X.~Tan, B.~Li, Y.~Liu, T.~Qin, sheng zhao, and T.-Y. Liu, ``Adaspeech: Adaptive text to speech for custom voice,'' in \emph{International Conference on Learning Representations}, 2021.

\bibitem{adaspeech2}
Y.~Yan, X.~Tan, B.~Li, T.~Qin, S.~Zhao, Y.~Shen, and T.-Y. Liu, ``Adaspeech 2: Adaptive text to speech with untranscribed data,'' in \emph{ICASSP 2021 - 2021 IEEE International Conference on Acoustics, Speech and Signal Processing (ICASSP)}, 2021, pp. 6613--6617.

\bibitem{hsieh23_interspeech}
C.-P. Hsieh, S.~Ghosh, and B.~Ginsburg, ``Adapter-based extension of multi-speaker text-to-speech model for new speakers,'' in \emph{Interspeech}, 2023, pp. 3028--3032.

\bibitem{UnitSpeech}
H.~Kim, S.~Kim, J.~Yeom, and S.~Yoon, ``Unitspeech: Speaker-adaptive speech synthesis with untranscribed data,'' in \emph{Interspeech}, 2023, pp. 3038--3042.

\bibitem{LoRA}
E.~J. Hu, yelong shen, P.~Wallis, Z.~Allen-Zhu, Y.~Li, S.~Wang, L.~Wang, and W.~Chen, ``Lo{RA}: Low-rank adaptation of large language models,'' in \emph{International Conference on Learning Representations}, 2022.

\bibitem{VoiceTailor}
H.~Kim, S.~gil Lee, J.~Yeom, C.~H. Lee, S.~Kim, and S.~Yoon, ``Voicetailor: Lightweight plug-in adapter for diffusion-based personalized text-to-speech,'' in \emph{Interspeech}, 2024, pp. 4413--4417.

\bibitem{librispeech}
V.~Panayotov, G.~Chen, D.~Povey, and S.~Khudanpur, ``Librispeech: An asr corpus based on public domain audio books,'' in \emph{2015 IEEE International Conference on Acoustics, Speech and Signal Processing (ICASSP)}, 2015, pp. 5206--5210.

\bibitem{ljspeech}
K.~Ito and L.~Johnson, ``The lj speech dataset,'' \url{https://keithito.com/LJ-Speech-Dataset/}, 2017.

\bibitem{zen19_interspeech}
H.~Zen, V.~Dang, R.~Clark, Y.~Zhang, R.~J. Weiss, Y.~Jia, Z.~Chen, and Y.~Wu, ``{LibriTTS: A Corpus Derived from LibriSpeech for Text-to-Speech},'' in \emph{Proc. Interspeech}, 2019, pp. 1526--1530.

\bibitem{autoguidance}
T.~Karras, M.~Aittala, T.~Kynk{\"a}{\"a}nniemi, J.~Lehtinen, T.~Aila, and S.~Laine, ``Guiding a diffusion model with a bad version of itself,'' \emph{arXiv preprint arXiv:2406.02507}, 2024.

\bibitem{chen21o_interspeech}
G.~Chen, S.~Chai, G.-B. Wang, J.~Du, W.-Q. Zhang, C.~Weng, D.~Su, D.~Povey, J.~Trmal, J.~Zhang, M.~Jin, S.~Khudanpur, S.~Watanabe, S.~Zhao, W.~Zou, X.~Li, X.~Yao, Y.~Wang, Z.~You, and Z.~Yan, ``Gigaspeech: An evolving, multi-domain asr corpus with 10,000 hours of transcribed audio,'' in \emph{Interspeech}, 2021, pp. 3670--3674.

\bibitem{Yamagishi2019-pe}
J.~Yamagishi, C.~Veaux, and K.~MacDonald, ``{CSTR} {VCTK} corpus: English multi-speaker corpus for {CSTR} voice cloning toolkit (version 0.92),'' 2019.

\bibitem{Interval}
T.~Kynk{\"a}{\"a}nniemi, M.~Aittala, T.~Karras, S.~Laine, T.~Aila, and J.~Lehtinen, ``Applying guidance in a limited interval improves sample and distribution quality in diffusion models,'' \emph{arXiv preprint arXiv:2404.07724}, 2024.

\bibitem{Adam}
D.~Kingma and J.~Ba, ``Adam: A method for stochastic optimization,'' in \emph{International Conference on Learning Representations (ICLR)}, San Diega, CA, USA, 2015.

\bibitem{gulati20_interspeech}
A.~Gulati, J.~Qin, C.-C. Chiu, N.~Parmar, Y.~Zhang, J.~Yu, W.~Han, S.~Wang, Z.~Zhang, Y.~Wu, and R.~Pang, ``{Conformer: Convolution-augmented Transformer for Speech Recognition},'' in \emph{Proc. Interspeech}, 2020, pp. 5036--5040.

\bibitem{kuchaiev2019nemo}
O.~Kuchaiev, J.~Li, H.~Nguyen, O.~Hrinchuk, R.~Leary, B.~Ginsburg, S.~Kriman, S.~Beliaev, V.~Lavrukhin, J.~Cook \emph{et~al.}, ``Nemo: a toolkit for building ai applications using neural modules,'' \emph{arXiv preprint arXiv:1909.09577}, 2019.

\bibitem{resemblyzer}
G.~Louppe, ``Resemblyzer,'' \url{https://github.com/resemble-ai/Resemblyzer}, 2019.

\bibitem{casanova24_interspeech}
E.~Casanova, K.~Davis, E.~Gölge, G.~Göknar, I.~Gulea, L.~Hart, A.~Aljafari, J.~Meyer, R.~Morais, S.~Olayemi, and J.~Weber, ``Xtts: a massively multilingual zero-shot text-to-speech model,'' in \emph{Interspeech}, 2024, pp. 4978--4982.

\bibitem{du2024cosyvoicescalablemultilingualzeroshot}
Z.~Du, Q.~Chen, S.~Zhang, K.~Hu, H.~Lu, Y.~Yang, H.~Hu, S.~Zheng, Y.~Gu, Z.~Ma, Z.~Gao, and Z.~Yan, ``Cosyvoice: A scalable multilingual zero-shot text-to-speech synthesizer based on supervised semantic tokens,'' 2024.

\bibitem{lee2023bigvgan}
S.~gil Lee, W.~Ping, B.~Ginsburg, B.~Catanzaro, and S.~Yoon, ``Big{VGAN}: A universal neural vocoder with large-scale training,'' in \emph{The Eleventh International Conference on Learning Representations}, 2023.

\end{thebibliography}

\end{document}